\documentclass[twocolumn,prl]{revtex4}

\usepackage{graphicx}
\usepackage{dcolumn}
\usepackage{amsmath}
\usepackage{bm}

\begin{document}

\title{Nonadiabatic population transfer in a tangent-pulse driven quantum model}
\author{Guang Yang}
\affiliation{Center of Theoretical Physics, College of Physical Science and
Technology, Sichuan University, Chengdu 610065, China}
\author{Wei Li}
\affiliation{Center of Theoretical Physics, College of Physical Science and
Technology, Sichuan University, Chengdu 610065, China}
\author{Li-Xiang Cen}
\email{lixiangcen@scu.edu.cn}
\affiliation{Center of Theoretical Physics, College of Physical Science and
Technology, Sichuan University, Chengdu 610065, China}

\begin{abstract}
Fine control of the dynamics of a quantum system
is the key element to perform quantum information processing and
coherent manipulations for atomic and molecular systems.
In this paper we propose a control protocol using a tangent-pulse
driven model and demonstrate that it indicates a desirable design,
i.e., of being both fast and accurate
for population transfer.
As opposed to other existing strategies,
a remarkable character of the present scheme is that high velocity of
the nonadiabatic evolution itself not only will not lead to unwanted
transitions but also can suppress the error caused by the truncation
of the driving pulse.
\end{abstract}

\maketitle

Dynamical control of quantum systems that undergo avoided level
crossings plays an important role in many areas of physics as well as some
of quantum chemistry \cite{book1,book2}. A well-known paradigm is that of
the Landau-Zener (LZ) model \cite{LZ1,LZ2} and its multi-state
extensions \cite{multi1,multi2,multi3,multi4,multi5,multi6,multi7}
which describe the evolution of quantum states in
the presence of a linearly changed external field. Owing to its very form of
the simplest driving field, the LZ model has become one of the mostly
investigated explicitly time-dependent quantum systems and is exploited as a
tool for controlling the population in various physical systems, e.g., optical
systems \cite{opt1,opt2,opt3,opt4}, semiconductor quantum dots \cite
{qdot1,qdot2,qdot3}, superconducting qubits \cite{superc1,superc2,superc3,superc4},
and so on.

The standard LZ model is defined in an infinite time domain
and the general solution of it, e.g., of the two-level case,
is described by the Weber's parabolic cylinder functions \cite{LZ1,LZ2}.
In the case of ideal driving, complete population transfer could be achieved
through the adiabatic passage of the linear LZ sweep.
As well as in other analogs of the LZ protocol,
population transfer
via avoided level crossings possesses the advantage of being insensitive
to the pulse area in comparison with the usual resonant $\pi$-pulse
scheme \cite{resonant}.
On the other hand, as the adiabatic evolution
is often required in these protocols, it indicates a slow speed---this
is an issue related to the generic quandary of quantum manipulations, i.e.,
how they could be implemented accurately and rapidly. Various strategies have
been proposed to tackle this issue, such as the composite adiabatic passage
technique \cite{CAP1,CAP2} and transitionless quantum driving in terms
of the counter-diabatic protocol \cite{cd1,cd2,cd3,cd4,note}
or the short-cut protocol \cite{shc1,shc2,shc3}.
In most of these cases, a complicated design of the driving field, e.g.,
in the transitionless driving algorithm an auxiliary counter-diabatic
field of time-dependent form, is required to achieve the
high-fidelity population transfer.

In this paper we propose a tangent-pulse driven quantum model
for nonadiabatic population transfer and demonstrate that,
conditioned to a matching sweep frequency, the transition dynamics
of the system, including the two-level case as well as its
multi-level extension,
is fully controllable in an analytical manner.
Contrary to those protocols based on the transitionless driving algorithm,
no auxiliary field is needed in the scheme and the nonadiabatic dynamics
of the model itself that undergoes avoided level crossings
could realize complete population transfer.
Not only that, but for the imperfect pulsing process with truncation,
we show that the high velocity of the nonadiabatic
evolution of the proposed model can suppress the error caused by the cutoff of the driving field.
As the protocol involves only a matching condition about the fixed sweep frequency,
it stands for an ideal design for fast and accurate population control
and may substitute the LZ model for potential applications.

The model we are considering is described by the following Hamiltonian
\begin{equation}
H(t)\equiv\mathbf{\Omega }(t)\cdot \mathbf{J}=\eta _1J_x+\eta _2\tan (\gamma
t)J_z,  \label{hamil}
\end{equation}
where $\mathbf{J}$ denotes the angular-momentum operator with the components
satisfying $[J_i,J_j]=i\varepsilon _{ijk}J_k$. The $z$ component of
the external field $\Omega _z(t)$ assumes a tangent-shape form [see Fig. 1(a)]
with $t\in(-\frac \pi {2\gamma },\frac \pi {2\gamma })$, and $\eta_{1,2}$ and
the sweep frequency $\gamma$ are fixed constants which satisfy the matching condition
(setting $\hbar =1$)
\begin{equation}
\eta_1^2=\eta_2^2+\gamma^2.
\end{equation}
Given a general tangent-shape pulse, e.g., $\tilde{
\mathbf{\Omega }}(t)=( \tilde{ \eta} _1,0, \tilde{\eta}_2\tan (\gamma t))$,
the above condition for the frequency could be fulfilled either by
tuning the $x$ component so that
$\tilde{\eta}_1\rightarrow \eta_1=\sqrt{\tilde{\eta }_2^2+\gamma ^2}$,
or by an overall modulation on the intensity
of the field, $\mathbf{\Omega }(t)\equiv \gamma \tilde{\mathbf{\Omega }}(t)/
\sqrt{\tilde{ \eta}_1^2-\tilde{\eta}_2^2}$, provided that $\tilde{\eta}_1>
\tilde{\eta}_2$.

Firstly we show that the dynamics of the system governed by the
Schr\"{o}dinger equation
\begin{equation}
i\frac \partial {\partial t}|\psi (t)\rangle =H(t)|\psi (t)\rangle
\label{schro}
\end{equation}
could be solved analytically. To this goal, we invoke a time-dependent
transformation on the wavefunction: $|\psi (t)\rangle =G(t)|\psi
^g(t)\rangle $, where $G(t)=e^{i\varphi J_z}e^{i(\gamma t+\pi /2)J_y}$ with
$\varphi =-\arcsin \frac {\gamma} {\eta _1}$. In the rotating frame defined
by $G(t)$, the state $|\psi ^g(t)\rangle $ satisfies a new Schr\"{o}dinger
equation, $i\partial _t|\psi ^g(t)\rangle =H^g(t)|\psi ^g(t)\rangle $, in
which the effective Hamiltonian $H^g(t)$ is obtained as
\begin{eqnarray}
H^g(t) &=&G^{\dagger }(t)H(t)G(t)-iG^{\dagger }(t)\partial _tG(t)  \nonumber
\\
&=&-\eta _2\cos ^{-1}(\gamma t)J_z.  \label{effham}
\end{eqnarray}
This simple form of $H^g(t)$, i.e., containing only the Cartan generator
$J_z$ but with vanishing other generators, is what one usually
expects in the algebraic approach to this kind of time-dependent quantum
systems. The previous examples successfully coped by the
algebraic method mostly involve driving fields with coordinately time-varying
components \cite{wang,alge,cen}.
The present system is distinctly different since
there is only one component depending on time and the corresponding
sweep generates a typical transition dynamics with avoided level crossings
which is in close analogy to that of the LZ model.

\begin{figure}[t]
\includegraphics[width=0.85\columnwidth]{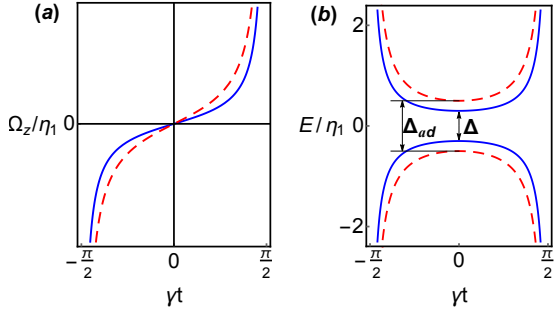}
\caption{Adiabatic versus nonadiabatic scanning processes of the tangent-pulse driven
model (\ref{hamil}) with a matching sweep
frequency: (a) Comparison of the time dependence of the $z$-component of the driving field
between the adiabatic sweep with $\gamma/\eta_1\rightarrow 0$ (dashed line)
and the nonadiabatic sweep with $\gamma/\eta_1=0.8$ (solid line);
(b) The corresponding adiabatic (dashed line) and diabatic (solid line)
energy levels, $E^{ad}_{\pm}(t)$ and $E_{\pm}(t)$ over the constant $\eta_1$
of the model with $j=\frac 12$, by which the associated gaps at the crossing point $t=0$
are shown to be $\Delta_{ad}=1$ and $\Delta=\eta_2/\eta_1=0.6$, respectively.}
\end{figure}

To proceed, noticing that $H^g(t)$ is Abelian along the time, the time
evolution of the system in the rotating frame is described by the
operator
\begin{equation}
U^g(t,t_0)=\exp [-i\int_{t_0}^tH^g(t^\prime)dt^\prime]=\exp [i\Theta
(t,t_0)J_z],  \label{timeo}
\end{equation}
where $\Theta (t,t_0)=\eta _2\int_{t_0}^t\cos ^{-1}(\gamma t^\prime)dt^\prime$.
That is to say, the Schr\"{o}dinger equation in this representation
possesses a stationary-state solution $|\psi _m^g(t)\rangle =e^{im\Theta
(t,t_0)}|m\rangle $ in which $|m\rangle $ denotes the eigenstate of $J_z$
with magnetic quantum number $m$. Hence the time evolution operator of the
original Schr\"{o}dinger equation (\ref{schro}) is given by
\begin{equation}
U(t,t_0)=G(t)U^g(t,t_0)G^{\dagger }(t_0);  \label{timeoo}
\end{equation}
and the basic solution to its wavefunction, the so-called diabatic base, is
obtained as
\begin{eqnarray}
|\psi _m(t)\rangle &=&G(t)|\psi _m^g(t)\rangle  \nonumber \\
&=&e^{im\Theta (t,t_0)}\sum_{m^{\prime }}\mathcal{D}_{m^{\prime }m}^j(\gamma
t+\frac \pi 2)e^{im^{\prime }\varphi }|m^{\prime }\rangle ,  \label{state}
\end{eqnarray}
where $\mathcal{D}_{m^{\prime }m}^j(\phi )\equiv \langle m^{\prime
}|e^{i\phi J_y}|m\rangle $ and the index $m^{\prime }$ of the summation is
taken over $-j,-j+1,\cdots ,j$ with $j$ the azimuthal quantum number.
Subsequently, the diabatic energy levels of the system can be calculated
according to $E_m(t)\equiv\langle \psi _m(t)|H(t)|\psi_m(t)\rangle $.

The result described above is applicable to the general SU(2) system, i.e.,
to the angular momentum with arbitrary quantum number $j$. For the simplest
two-level ($j=\frac 12$) system, there is
\begin{equation}
\mathcal{D}^{\frac 12}(\phi )=\left(
\begin{array}{ll}
\cos \frac \phi 2 & \sin \frac \phi 2 \\
-\sin \frac \phi 2 & \cos \frac \phi 2
\end{array}
\right) ,  \label{dtwo}
\end{equation}
so the diabatic bases are shown to be
\begin{equation}
|\psi _{\pm \frac 12}(t)\rangle =e^{\pm \frac i2\Theta (t,t_0)}\left(
\begin{array}{l}
\cos (\frac{\gamma t}2\pm \frac \pi 4)e^{i\varphi /2} \\
-\sin (\frac{\gamma t}2\pm \frac \pi 4)e^{-i\varphi /2}
\end{array}
\right) .  \label{basistwo}
\end{equation}
The corresponding diabatic energy levels are obtained as
$E_{\pm }(t)=\mp \frac 12\eta_2\cos^{-1}(\gamma t)$, which together
with the adiabatic
$E^{ad}_{\pm}(t)=\mp \frac 12\eta_1\cos^{-1}(\gamma t)$,
are depicted in Fig. 1(b).
It is seen that at the
beginning and the ending points of the sweep,
the diabatic basis state $|\psi _{\frac
12}(t)\rangle $ tends to $|+\rangle =$( {\tiny $\!
\begin{array}{l}
1 \\
0
\end{array}
$}) as $t\rightarrow -\frac \pi {2\gamma }$ and to $|-\rangle =$({\tiny $\!
\begin{array}{l}
0 \\
1
\end{array}
$}) as $t\rightarrow +\frac \pi {2\gamma }$ (up to a phase term); and the
basis state $|\psi _{-\frac 12}(t)\rangle $ tends reciprocally to $|-\rangle
$ or $|+\rangle $ as $t\rightarrow \mp \frac \pi {2\gamma }$.
Therefore the complete
population transfer $|+\rangle \leftrightarrow |-\rangle $ could be realized
through the ideal tangent-pulse sweep, whatever the driving process is
adiabatic or nonadiabatic.

\begin{figure}[b]
\includegraphics[width=0.85\columnwidth]{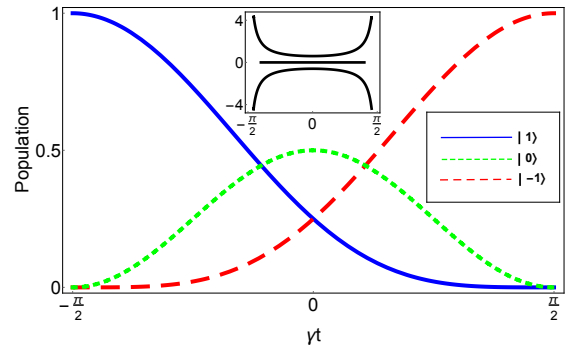}
\caption{Nonadiabatic population transfer along the time evolution
in the tangent-pulse driven model (\ref{hamil}) with $j=1$.
The initial state is in $|1\rangle$. The inset describes the
diabatic levels $E_m(t)/\eta_1$ ($m=1,0,-1$) with $\gamma/\eta_1=0.8$
and they exhibit the avoided crossing at $t=0$.}
\end{figure}

The Hamiltonian (\ref{hamil}) with high quantum number $j$ accounts for a
multi-level extension of the sweep protocol associated with
avoided crossings. In particular, the exact
solvability of the system makes it an ideal scenario to manifest
the behavior of the wavefunction undergoing
multichannel transitions. Firstly, it is direct to show that the
driving process results in a transition from the state $|m\rangle $ to
$|-m\rangle $ for all possible $m=-j,-j+1,\cdots ,j$. According to
Eq. (\ref{timeoo}), the evolution operator generated by the overall
sweep during $t\in (-\tau,\tau)$ with $\tau\rightarrow \frac
{\pi} {2\gamma}$ takes the form of
\begin{equation}
U(\tau,-\tau)=e^{i\varphi J_z}e^{i\pi J_y}e^{i[\Theta(\tau)-\varphi]J_z},
\end{equation}
in which $\Theta (\tau)=\eta_2\int_{-\tau}^{\tau}\cos^{-1}(\gamma t)dt$.
In view of the transformation $e^{i\pi J_y}|m\rangle \rightarrow |-m\rangle$,
the transition probability between any two states
$|m\rangle$ and $|m^\prime \rangle $ is immediately yielded: $|\langle
m^\prime|U(\tau,-\tau)|m\rangle |^2=\delta _{m^\prime,-m}$.
To illustrate further the multichannel transition process
of the wavefunction, we resort to the three-level case with $j=1$.
The corresponding matrix $\mathcal{D}_{m^{\prime }m}^j(\phi )$ is shown as
\begin{equation}
\mathcal{D}^1(\phi )=\left(
\begin{array}{lll}
\cos ^2\frac \phi 2 & \frac{\sin \phi }{\sqrt{2}} & \sin ^2\frac \phi 2 \\
-\frac{\sin \phi }{\sqrt{2}} & \cos \phi & \frac{\sin \phi }{\sqrt{2}} \\
\sin ^2\frac \phi 2 & -\frac{\sin \phi }{\sqrt{2}} & \cos ^2\frac \phi 2
\end{array}
\right) .  \label{dthree}
\end{equation}
By taking $\phi=\gamma t+\frac \pi 2$ and substituting the above expression
into Eq. (\ref{state}), one obtains the diabatic bases for
the three-level system
\begin{widetext}
\begin{eqnarray}
|\psi _0(t)\rangle &=&\frac {\sqrt{2}}{2}\cos (\gamma t)e^{i\varphi }|1\rangle
-\sin (\gamma t)|0\rangle -\frac {\sqrt{2}}{2}\cos (\gamma t)e^{-i\varphi
}|-1\rangle ,
\nonumber
\\
|\psi _{\pm 1}(t)\rangle &=&e^{\pm i\Theta (t,t_0)}\{\frac 12[1\mp \sin(\gamma t)]e^{i\varphi
}|1\rangle \mp\frac {\sqrt{2}}{2}\cos(\gamma t)|0\rangle +\frac 12[1\pm\sin (\gamma t)]
e^{-i\varphi }|-1\rangle \}.
\label{three}
\end{eqnarray}
\end{widetext}
Starting from an initial state $|\psi(-\tau)\rangle =|1\rangle$,
the wavefunction will undergo transitions via the channels
$|1\rangle \rightarrow |-1\rangle$ and $|1\rangle \rightarrow |0\rangle
\rightarrow |-1\rangle$, and evolve precisely to the ending state $|-1\rangle$
up to a phase factor. We depict in Fig. 2 the corresponding population
transfer along the time evolution of the driving process.
It is seen that the maximal population in the
intermediate state $|0\rangle $ is $p=\frac 12$ which occurs at the point
$t=0$.

In the practical scanning process, the driving field has finite intensity
and the cutoff of the sweep will cause some losses of the transition
probability. Consider that the truncation of the field pulse is symmetric.
For the evolution generated during the period $t\in[-\tau_c,\tau_c]$, the
goal now is to find the matrix of transition probabilities $\hat{P}(\tau_c)$,
with the matrix element defined as $P_{m^{\prime}m}(\tau_c)\equiv|\langle
m^{\prime}|U(\tau_c,-\tau_c)|m\rangle|^2$. Denote by $\delta\equiv\frac
{\pi} {2}-\gamma \tau_c$ the deviation of the maximal phase
angle from that of the ideal tangent pulse. Since the evolution operator
$U(\tau_c,-\tau_c)$ is specified explicitly in Eq. (\ref{timeoo}),
$P_{m^{\prime}m}(\tau_c)$ could be calculated directly via
\begin{equation}
P_{m^{\prime}m}(\tau_c) =|\langle m^{\prime} |e^{i(\pi-\delta )J_y}e^{i\Theta
(\tau_c)J_z}e^{-i\delta J_y}|m\rangle |^2,
\label{pro1}
\end{equation}
in which $\Theta (\tau_c)\equiv \Theta (\tau_c,-\tau_c)$.
As the representatives $\mathcal{D}_{k^\prime k}^j(-\delta)=\langle
k^\prime|e^{-i\delta J_y}|k\rangle$ for $j=\frac 12$ and $j=1$
are already given in Eqs. (\ref{dtwo}) and (\ref{dthree}),
the corresponding matrices of the transition probabilities of these two
cases are readily obtained. In the following we shall focus on
the influence of the sweep frequency $\gamma$ on the
transition probability and demonstrate that
the high velocity of the scanning rate could suppress the error
caused by the truncation of the driving pulse.
We stress that this is a general result for the
described driven model, although it will be illustrated
below via the simplest case of $j=\frac 12$.

Explicitly, for the two-level system undergoing an imperfect
driving process with the symmetric truncation,
the probability of the transition
$|+\rangle \leftrightarrow |-\rangle$ is given by
\begin{equation}
P_{-+}=P_{+-}=1-\cos ^2\frac{\Theta
(\tau _c)}2\sin ^2 \delta.
\label{pro2}
\end{equation}
To reveal the influence of the sweep frequency $\gamma$ on the population transfer,
let us denote by $\delta_0=\arctan\frac{\Omega_x}{\Omega _z(\tau_c)}$ the
deviation of the pulsed field vector $\mathbf{\Omega}(t)$ from
the ideal $z$ axis at the points $t=\pm\tau_c$.
In view of $\Omega_x=\eta_1$, $\Omega_z(\tau_c)=\eta_2\tan(\gamma \tau_c)$, and
the matching condition of the frequency: $\eta_2^2=\eta_1^2-\gamma^2$,
one has
\begin{equation}
\tan\delta=\sqrt{1-(\gamma/\eta_1)^2}\tan\delta_0.
\label{rela}
\end{equation}
The equations (\ref{pro2}) and (\ref{rela}) indicate that,
for the fixed values of $\Omega_x$ and $\Omega_z(\tau_c)$,
the cutoff error to the transition probability
that is dominated by $\delta$ could be suppressed
through increasing the sweep frequency $\gamma$.
Especially, as long as the frequency is modulated within the
matching condition and satisfies
$\gamma/\eta_1\rightarrow 1$,
the high-fidelity population transfer could be achieved
by the nonadiabtic evolution even when there exists dramatic truncation of
the driving field, i.e., with a finite deviation $\delta_0$ of the field vector
from the $z$ axis at $t=\pm \tau_c$ (see Fig. 3).

The above characterized dynamics of the tangent protocol in which the population
transfer could be enhanced via accelerating the scanning rate
implies a rare and intriguing character that has not ever been found
in other existing quantum driven models.
Firstly, in the linear LZ protocol the nonadiabaticity of
the evolution is known to induce unwanted transitions to
the population transfer. A second example appropriately
serving as a reference is the counter-diabatic protocol
based on the transitionless quantum driving.
Given a Hamiltonian $H(t)$, the protocol cancels
the nonadiabatic part of the evolution under
$H(t)$ by introducing an auxiliary counter-diabatic field
and ensures that the system evolving under the total Hamiltonian
$H_{cd}(t)=H(t)+H^{\prime }(t)$ always remains in the instantaneous
eigenstate of $H(t)$. For the Hamiltonian specified by Eq. (\ref{hamil}),
there is $H_{cd}(t)=H(t)+\dot{\delta}_{cd}(t)J_y$ and
the corresponding time evolution operator reads
\begin{equation}
U_{cd}(t,t_0)=e^{i\delta_{cd}(t)J_y}e^{i\Theta_{cd}(t,t_0)J_z}e^{-i\delta_{cd}
(t_0)J_y},
\label{counterh}
\end{equation}
in which $\delta_{cd}(t)=\pi-\arccos \frac{\Omega_z(t)}{\Omega (t)}$ and
$\Theta_{cd}(t,t_0)=\int_{t_0}^{t}\Omega (\tau)d\tau$ with
$\Omega (t)=\sqrt{\Omega_x^2+\Omega_z^2(t)}$.

\begin{figure}[t]
\includegraphics[width=0.85\columnwidth]{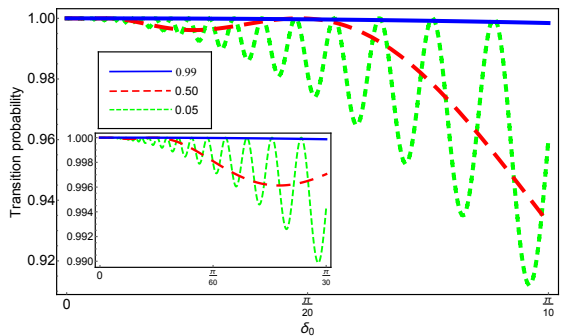}
\caption{Transition probabilities yielded by the imperfect pulsing processes
with truncation. The error caused by the cutoff of the driving field is
shown to be suppressed when the scanning rate of the protocol increases.
An ultrahigh fidelity ($1-P_{-+}\sim 10^{-4}$) is attainable through the
nonadiabatic evolution ($\gamma/\eta_1=0.99$) even
when there is dramatic truncation of the driving field
($\delta_0\sim \frac {\pi}{30}$).}
\end{figure}

Consider a similar truncation of the scanning process of
the counter-diabatic protocol, i.e.,
$t\in[-\tau_c,\tau_c]$ with $\gamma \tau_c=\frac {\pi}{2}-\delta$.
The matrix of transition probabilities related to the protocol
is specified accordingly as $P^{cd}_{m^\prime m}(\tau_c)\equiv
|\langle m^\prime|U_{cd}(\tau_c,-\tau_c)|m\rangle|^2$ in which
$U_{cd}(\tau_c,-\tau_c)$ denotes the corresponding evolution operator
described in Eq. (\ref{counterh}).
In view of $\delta_{cd}(-\tau_c)=\delta_0$
and $\delta_{cd}(\tau_c)=\pi-\delta_0$, it is readily recognized that
for the two-level system the transition
probability reads
\begin{eqnarray}
P^{cd}_{-+}(\tau_c)
&=&|\langle -|e^{i(\pi-\delta_0)J_y}e^{i\Theta
_{cd}(\tau_c)J_z}e^{-i\delta_0 J_y}|+\rangle|^2
\nonumber \\
&=&1-\cos^2\frac {\Theta_{cd}(\tau_c)}{2}\sin^2\delta_0,
\label{pro3}
\end{eqnarray}
in which $\Theta_{cd}(\tau_c)\equiv\Theta_{cd}(\tau_c,-\tau_c)$.
Comparing with Eq. (\ref{pro2}), the critical difference is that
the cutoff error in the counter-diabatic protocol is
dominated by $\delta_0$ instead of $\delta$ that was shown
in the previously described protocol.
Indeed, since the transitionless driving algorithm
pursues the evolution in such a way that the state
remains to be in the adiabatic state of $H(t)$,
it turns out to be a general result that the cutoff error in all
its resulting protocols is independent of the sweep frequency.

Summing up, we have proposed a design for the population control
which unifies the high operation rate and robustness as its
intrinsic character.
While its simple form of the tangent-pulse sweep could rival the
linear LZ protocol, we have shown that the generated dynamics
associated with avoided level crossings,
whatever adiabatic or nonadiabatic,
could yield the desired population transfer.
Compared to those existing nonadiabatic protocols based on the
transitionless driving algorithm, the present scheme possesses distinct
superiorities: 1) no auxiliary time-varying field but a simple
matching condition for the sweep frequency is involved;
2) for imperfect pulsing processes with truncation, the cutoff error
could be suppressed by enhancing the scanning rate of the protocol.
We emphasize that the matching condition of the fixed frequency
in the design does not add technical complexity to the tangent-pulse
driving and is readily achievable for experimental implementation,
e.g., by the Bose-Einstein condensates in an accelerated optical lattice
\cite{bec} or by the nitrogen-vacancy center in diamond
\cite{cd3}. Finally, we expect that the finding of the exactly solvable
tangent-pulse driven quantum model is helpful to advance
further the study of the issue of the solvability for more general
time-dependent quantum systems.


\begin{references}
\bibitem{note} The same idea that makes use of an additional interaction to eliminate
the nonadiabatic effect has ever been exploited in the scheme of the
holonomic quantum computation, see Ref. \cite{cen}.
\end{references}
\end{document}